\documentclass[sigconf]{acmart}

\AtBeginDocument{%
  \providecommand\BibTeX{{%
    \normalfont B\kern-0.5em{\scshape i\kern-0.25em b}\kern-0.8em\TeX}}}

\copyrightyear{2022}
\acmYear{2022}
\setcopyright{acmlicensed}\acmConference[FAccT '22]{2022 ACM Conference
on Fairness, Accountability, and Transparency}{June 21--24, 2022}{Seoul,
Republic of Korea}
\acmBooktitle{2022 ACM Conference on Fairness, Accountability, and
Transparency (FAccT '22), June 21--24, 2022, Seoul, Republic of Korea}
\acmPrice{15.00}
\acmDOI{10.1145/3531146.3533137}
\acmISBN{978-1-4503-9352-2/22/06}
    
\usepackage{etoolbox}
\usepackage{longtable}

\begin{document}

\title{Making the Unaccountable Internet: \\ The Changing Meaning of Accounting in the Early ARPANET}


\author{A. Feder Cooper}
\affiliation{%
   \institution{Cornell University}
   \department{Department of Computer Science}
   \city{Ithaca}
   \state{NY}
   \country{USA}}
   \email{afc78@cornell.edu}

%\authornote{The authors use alphabetical superscripts for footnotes; numerically super-scripted end notes can be found in the Appendix.}

\author{Gili Vidan}
\affiliation{%
   \institution{Cornell University}
   \department{Department of Information Science}
   \city{Ithaca}
   \state{NY}
   \country{USA}}
   \email{gv232@cornell.edu}

\renewcommand{\shortauthors}{Cooper and Vidan}
\renewcommand{\shorttitle}{Making the Unaccountable Internet}

\begin{abstract}
Contemporary concerns over the governance of technological systems often run up against narratives about the technical infeasibility of designing mechanisms for accountability. While in recent AI ethics literature these concerns have been deliberated predominantly in relation to machine learning, other instances in the history of computing also presented circumstances in which computer scientists needed to un-muddle what it means to design accountable systems. One such compelling narrative can frequently be found in canonical histories of the Internet that highlight how its original designers' commitment to the ``End-to-End'' architectural principle precluded other features from being implemented, resulting in the fast-growing, generative, but ultimately unaccountable network we have today. This paper offers a critique of such technologically essentialist notions of accountability and the characterization of the ``unaccountable Internet'' as an unintended consequence. It explores the changing meaning of accounting and its relationship to accountability in a selected corpus of requests for comments (RFCs) concerning the early Internet's design from the 1970s and 80s. We characterize four ways of conceptualizing accounting: as \emph{billing}, as \emph{measurement}, as \emph{management}, and as \emph{policy}, and demonstrate how an understanding of accountability was constituted through these shifting meanings. We link together the administrative and technical mechanisms of \textit{accounting} for shared resources in a distributed system and an emerging notion of \textit{accountability} as a social, political, and technical category, arguing that the former is constitutive of the latter. Recovering this history is not only important for understanding the processes that shaped the Internet, but also serves as a starting point for unpacking the complicated political choices that are involved in designing accountability mechanisms for other technological systems today.
\end{abstract}




\maketitle

\vspace{-.2cm}
\section{Introduction} \label{sec:intro}

There has recently been a revival of interest in the value of accountability in technical systems, especially those that involve machine learning. %Early on, 
Already in 1996, \citet{nissenbaum1996accountability} recognized the unique challenges of accountability in a computerized society; from a philosophical perspective, she discussed how the ascendance of %computer technology's use 
computerization in important societal contexts was poised to introduce novel barriers to accountability. 25 years later, one can observe the difficulties~\citet{nissenbaum1996accountability} raised playing out in recent AI ethics scholarship, illustrating that accountability is a slippery term with multiple valences~\cite{cooper2022accountability}. For example, several scholars have imported  a definition of ``public accountability'' from~\citet{bovens2007accountability} into research concerning algorithmic accountability~\cite{wieringa2020accountability, kacianka2021accountable}, while others have attempted to institute cultures of accountability through \emph{ex ante} engineering standards of care~\cite{gebru2018datasheets,mitchell2019modelcards} or \emph{post hoc} audits and impact assessments~\cite{raji2020audit, vecchione2021audit, adler2018auditing, metcalf2021aia, moss2021assembling}. These works draw from philosophy, computer science, and the law, and focus almost exclusively on accountability in relation to \emph{algorithms}. This emphasis on algorithms has left broader systems and institutions%al concerns
 (in which algorithms are situated) unattended in terms of their relationship to accountability~\cite{cooper2021eaamo}.

In this paper, we provide a complement to prior conceptual technological accountability scholarship. We take a historical approach and examine accountability as it pertains to a specific computer system, situated in a time and a place, including its architectural components, protocols, and institutional actors. Rather than taking a top-down approach in which we posit our own definition of accountability to ground our analysis~\cite{wieringa2020accountability, kim2021accountability, kroll2017aa, kroll2021traceability}, we investigate how an understanding of unaccountability can emerge and evolve over time during a system's development. For our subject, we concentrate on the early Internet and its predecessor---the ARPA-controlled ARPANET.\footnote{ARPA, the Advanced Research Projects Agency, has gone%see-sawed
back and forth on including a D for ``Defense'' at the start of its name. The agency settled on DARPA in 1996, %(since 1996), 
but we refer to it as ARPA throughout for clarity, since our subject, the ARPANET (sometimes referred to as the ARPAnet), was named after ARPA~\cite[p. 103]{leiner1997pastandfuture}~\cite{clark2018internet}.} Today, the Internet is often described as lacking endemic features for accountability, fixing the perception it is an \emph{un}accountable system~\cite{rexford2010cleanslate, clark2018internet, weitzner2008infoaccount}. We contend that identifying how this perception developed can help locate what it means for a system to be accountable, and can enable us to interrogate the dynamics that lead a research community to develop an unaccountable system. While our work is a deep dive into the particular origins of the unaccountable Internet, it is also a case study that elicits broader lessons concerning the unaccountability of complex computer systems, and thus serves as a cautionary tale for the design of contemporary systems.

\subsection{Retelling Internet Histories: From End-to-End to Accounting}\label{sec:intro:endtoend}

Over the past two decades, Internet historians have pointed to the institutional context of the network's protocols and standards' development~\cite{abbate1999inventing,braman2013governance,russell2014open,yates2019rules, denardis2015surveillance}, explored how commercial actors, user groups, and regulators also shaped its trajectory~\cite{turner2006culture, greenstein2015internet,mcilwain2020software}, and contrasted the US-origins of the Internet with networks developed in different political contexts~\cite{medina2011chile, peters2008soviet}. We follow these studies by looking at how network engineers and administrators, as a research community and governing institution, defined the goal of networking and often contrasted it with changing meanings of %the concepts of 
``accounting'' and ``accountability.'' This paper surveys %both 
the %very 
early years of the ARPANET project, what \citet{braman2011framing} has %referred to as 
called the ``framing years,'' and goes through the transition of the ARPANET to the Internet. During this decades-long period, the governing bodies and structures of the networks in question were changing, even if many of the same individuals developing network computing remained involved. %several the decades. 
We therefore refer to this community as ``network engineers'' or ARPANET contractors in its early years, %as well 
and as its more formalized afterlives---%as 
the Internet Activities Board (IAB) and the Internet Engineering Task Force (IETF).\footnote{\citet[pp. 50-53]{russell2006war} provides a brief but insightful overview of the process of expansion and formalization of the network research and development community that this paper examines.}

These institutional perspectives on the history and politics of networks contrast with other accounts that focus on the architectural novelty of the network's design. At the center of these architectural histories stands the now foundational ``End-to-End principle.'' In the early 1980s, Jerome Salzer, Dave Reed, and David Clark, three members of the ARPANET project team, formalized the End-to-End principle in a paper, arguing that a communication network should implement most complex functions at the endpoints rather than within the network itself, leaving the network's design relatively simple and focused on routing data traffic~\cite[p. 278]{saltzer1984endtoend}.\footnote{The principle %had been tacitly 
was a tacit %working 
assumption of the ARPANET project (among other %computing 
research efforts) for over a decade by the time the paper made it explicit~\cite[p. 49]{russell2006war}~\cite{saltzer1984endtoend}.} Through various retellings of the Internet's origin myth, End-to-End became its defining feature, used to explain the network's exponential, rapid growth and celebrated for its architectural simplicity and even its political commitment to decentralization, individual autonomy, and freedom---and later, a reason behind some of the Internet's discontents~\cite{blumenthal2001endtoend, rexford2010cleanslate}.\footnote{Legal scholars attempting to theorize the relationship between network architecture and political order in particular zeroed in on End-to-End as a way to tie the two together~\cite{lessig2006code2,greenstein2015internet,zittrain2008book, laundau2011privacy, post2009jefferson}.} The ubiquity of End-to-End as a stand-in for the Internet as a whole was noted by Tarleton Gillespie, who argued that such a cultural uptake of one specific term both obfuscated its contentious meaning and created a sense in which the architecture of the Internet became a fixed material object~\cite{gillespie2006endtoend}.\footnote{This cultural uptake had broader reverberations among users of the network. In November 1988, the Morris Worm caused failures that brought down the majority of nodes on the Internet~\cite{rfc1135}. The network seemed to be in crisis. That such widespread failure could occur suggested the existing, End-to-End-guided architecture was perhaps not suited to the network's broader use and needs. (A retrospective analysis of the Worm is out of the scope of this paper. Please refer to~\citet{slayton2016worm}.) Nevertheless, the culture of End-to-End prevailed. After the initial response to excise the Worm from the network, what followed shortly after was a series of ethics statements from revered institutions, including the IAB, MIT, and the NSF~\cite{nsf1989code,mit1989code,cpsr1989code, iab1989code, rfc1087}. These statements  centered the \emph{end}-user as the site at which to locate the responsibility for appropriate network use, rather than calling for proposals to imagine what designing for accountability could mean for internals of the network.} This End-to-End ethos has arguably carried forward to the present more generally in computing, with \citet{zittrain2014blog} recasting End-to-End as the ``procrastination principle'' to capture the idea that ``most problems could be solved later and by others.''\footnote{Zittrain discusses the principle in terms of the network in this blog post, and discusses its broader import for software engineering in his book~\cite{zittrain2008book}.} One can see this in several places, for example: invoked in Google’s now-retired motto, ``Don’t be evil,’’ a rhetorical move that could be read as punting accountability for the behavior of its products to individual engineers; in evolving conversations concerning balancing accountability and unconstrained user behavior with respect to platform content moderation~\cite{klonick2020facebook, gillespie2010platform, kosseff2022sec230, citron2022privacy}; and in corporate and institutional ethics statements targeted at individual engineers to use machine learning and artificial intelligence technology responsibly.

Among the many possible narratives that get lost in the End-to-End telling is the story of a difficult administrative, technical, economic, and political problem at the heart of the ARPANET: distributed accounting. In the late 1980s, David Clark, one of the trio behind the original 1980s End-to-End paper, reflected back on the experience of designing the ARPANET and Internet protocols and described a set of goals that guided the design principles of the network. The top-level goal was to create a means of interconnecting existing computer networks, supporting the connection of different devices and pre-existing communication protocols~\cite[pp. 1-2]{clark1988darpa}. It is in relation to this top-level goal that End-to-End was often celebrated as an elegant solution. But Clark went on to articulate a list of seven ``secondary goals,'' last among which read: ``resources used on the network must be accountable''~\cite[pp. 1-2]{clark1988darpa}. Clark refers to this as the goal of ``accountability'' and cites earlier work on the ARPANET by Vint Cerf and Bob Kahn that already noted the need for a network to provide such features in 1974. The %1974 Cerf and Kahn 
paper in question, however, only makes mention of ``accounting'' for traffic in relation to the potential to charge for network use~\cite[p. 2]{cerfkahn1974tcp}. 

In this paper, we propose to take seriously the slippage between Clark's 1988 use of ``accountability'' and the subordinate, discounted goal of accounting for resources. We argue that matters of accounting were never \emph{merely} about charging and billing, nor were they straightforward---they raised a variety of questions that form a theory of accountability in a networked environment. We trace how the historical actors designing the technical system that is the Internet routinely debated, deferred, and discounted accounting as a feature of the network, and we see how, over time, this community came to define the meaning of accounting as constitutive of accountability.\footnote{Daniel Neyland has argued that in the context of machine learning algorithms, there is an intersection between the two registers of the term ``account,'' suggesting that calls to make algorithms more accountable should take note of system features that make technical systems account-able~\cite{neyland2016accountable}. We similarly focus on the intersection of these two registers, but, by taking a historical approach, we go beyond suggesting a connection to tracing its development.} In tracing the shifting meaning of accounting in the first two decades of network computing, we offer a retelling of Internet histories, locating the unresolved problem of accounting as the obverse of the End-to-End elegant solution story.

\subsection{Research Method and Contribution}\label{sec:intro:method}

We enlist a variety of secondary sources, including %books and papers 
work by historians and Science and Technology Studies (STS) scholars. Our main focus is primary sources drawn from the self-described ``old boys network''~\cite[p. 54]{abbate1999inventing} that participated in the architecture discussions and implementation of the ARPANET, the Internet’s US predecessor during the 1970s. We %then 
follow this community's writings through the 1980s---the early days of the Internet---bringing in  %These sources include 
contemporary retrospectives of Internet history written by networking researchers like David Clark, as well as early networking conference proceedings, journal articles, and technical reports. However, we primarily focus on a corpus of Internet Requests for Comments (RFCs): a chronologically-ordered series of documents posted publicly and hosted online by the Internet Society. %As \citet{braman2013governance} argues, 
The RFCs present multiple histories and can be read for their technical content, a community's self-reflection on its own history, and as source materials for analysts of governance formation~\cite[p. 66]{braman2013governance}. The status and means of distributing the RFCs was also subject to change over the period we examine.\footnote{Originally, the RFCs were necessarily circulated on paper through the physical mail system, and then later were hosted at Stanford Research Institute (SRI)~\cite{abbate1999inventing}. The current \texttt{rfc-editor} tool is available at~\url{https://www.rfc-editor.org/}.} Our use of the RFCs was therefore a helpful way to map the contours of the community's conversations over accounting and accountability.\footnote{We, however, did not consider the RFCs to be the sole means through which such conversations took place (see Appendix~\ref{app:sec:methods}). For more on how media, legal, and policy scholars have approached reading RFCs see: \citet[pp. 40-43]{milne2021email} and \citet{braman2010interpenetration}.}

We scraped and filtered the entire corpus to find instances of RFCs that mention ``accounting'' and ``accountability.'’ Our process was not just a matter of collecting and counting all documents from the first RFC to mention ``accounting''~\cite[p. 3]{rfc33},\footnote{RFC 33 represents the first mention of accounting; it added a notion of accounting to the earlier host-host protocol, originally defined in 1969 in RFC 11~\cite{rfc11}.} to the first RFC that uses the word ``accountability''~\cite[p. 2]{rfc721}.\footnote{While RFC 721 in 1976 is the first to use the word ``accountability,'' arguably, it was RFC 808 in 1982, documenting  a January 10, 1979 meeting at BBN, which was the first to distinguish between ``accounting'' and ``accountability'' explicitly~\cite[p. 3]{rfc808}: ``There was some general discussion of the impact of personal computers on mail services.  The main realization being that the personal computer will not be available to handle incoming mail all the time.  Probably, personal computer users will have their mailboxes on some big brother computer (which may be dedicated to mailbox service, or be a general purpose host) and poll for their mail when they want to read it. \emph{There were some concerns raised about accountability and accounting}'' (emphasis added). } Rather, we performed a detailed reading to produce an understanding of how accountability evolved conceptually from accounting over time. %(see Appendix~\ref{app:sec:methods} for more details on our methods)
By our accounting, \emph{accounting} moved through being understood as \emph{billing} (Section~\ref{sec:billing}), as \emph{measurement} (Section~\ref{sec:measurement}), as \emph{management} (Section~\ref{sec:management}), and as \emph{policy} (Section~\ref{sec:policy}), a point at which \emph{accounting} was understood as constitutive of \emph{accountability}, and the lack of accounting features developed within the network's design an impediment to the network's operations. The rest of the paper is organized around these four notions, with some nonlinear, overlapping chronological progression. We begin with documents dating from before the first RFC---prior to the proposal of the ARPANET project~\cite{roberts1967arpanetproposal}---and trace the changing meaning of accounting and %to 
accountability through 1990, when the ARPANET was decommissioned~\cite[p. 195]{abbate1999inventing}~\cite{cerf1989requiem}. We scope our project to this period because, while concerns about accountability clearly extend through to present day discussions of Internet governance~\cite{klonick2020facebook, clark2018internet, citron2022privacy}, we found that it was during this time period that a notion of an unaccountable network first came about and took hold as a defining characterization of the Internet.

By identifying the changing meanings of accounting in the ARPANET and early Internet, we offer the following three contributions for the study of accountability in technical systems: first, we link together the administrative and technical mechanisms of \textit{accounting} for shared resources in a distributed system and an emerging notion of \textit{accountability} as a social, political, and technical category, arguing that the former is constitutive of the latter. Second, we characterize a research dynamic among the technical community we studied that deprioritizes the development of administrative tools and accounting infrastructure, treating it as existing somewhere beyond the scope of their work as a matter of ``policy'' or as subordinate to their core research objectives. Third, this retelling of the early history of the Internet offers not only a corrective for how we view its particular development, but also provides significant lessons about the role of having institutional structures in place and designing for and around their administrative needs when building new accountable technological systems.

\section{Accounting as billing: \\The double-bind of sharing networked resources} \label{sec:billing}

``ARPA will not pay for the coffee and pastry being served, so please chip in to help me pay for it''~\cite[p. 2]{rfc82}.  This was early-Internet engineer Steve Crocker's introductory remark to the Network Working Group (NWG) of the ARPANET project in its November 16, 1970 meeting in Houston, Texas. While this may seem like an inane detail about NWG bookkeeping, it in fact serves to highlight the moment when there began a major shift concerning who was paying for using ARPA resources, and what exactly constituted the use that needed to be paid for. That is, while ARPA's Information Processing Techniques Office (IPTO) had comprehensively funded its researcher-contractors' logistical and capital expenses since its inception in 1962, by 1970, IPTO-funded labs felt the purse strings tighten. Low-level, seemingly incidental operational costs like a working group's coffee bill warranted space in the official NWG meeting record. Amid the gravitas of what its members even at the time recognized would be pivotal discussions concerning the architecture of the first distributed computing network, the question of who pays for the coffee was never too far from the question of who pays for computing~\cite[p. 18]{rfc82}.\footnote{Starting in 1962, IPTO, situated within the broader ARPA umbrella, essentially funded computer science research centers across the US (including at MIT, UCLA, and Carnegie Mellon), ``often outspending universities significantly'' in terms of research support~\cite[pp. 36-37, 44, 56]{abbate1999inventing}. These research centers followed the local time-sharing computing paradigm~\cite{carr1970rfc33pub}. Also in 1962, J. C. R. Licklider and Welden E. Clark wrote their piece on ``On-line man-computer communication''~\cite{licklider1962network} that is frequently cited as the first work to discuss communicating, interconnected computers that can share data and programs~\cite{leiner2009briefhistory, leiner1997pastandfuture}. For more on the 1960s development of a notion of an interconnected computer network, both by Licklider and among other researchers, see~\citet[Ch. 1]{abbate1999inventing}, \citet{turner2006culture}, and \citet{aspray2008commercial}.}

Talking about the cost of resources was new for beneficiaries of IPTO's funding. Such operational details had not been a concern when there was no network through which computing resources were to be shared. IPTO had purchased computers for its contracted computing sites, which were working on local, site-specific projects, such as automated theorem proving at Stanford Research Institute (SRI) and natural language processing at Bolt, Beranek, and Newman (BBN)~\cite[p. 548]{robertswessler1970resourcesharing}~\cite[p. 44]{abbate1999inventing}.\footnote{IPTO/ARPA also leased communication lines from common carriers, to serve as the physical connection medium between remote nodes~\cite{roberts1967arpanetproposal}~\cite[pp. 167-8]{russell2014open}.} Having paid for these computers up front, IPTO was not particularly concerned about the low-level specifics of how they were used. From the perspective of IPTO-contract-site researchers, this hands-off policy enabled conditions of unrestricted, free usage; it seemed like contracting with ARPA was easy money, as the funding seemed to come with ``few strings attached''~\cite[p. 77]{abbate1999inventing}.

However, this status quo of unchecked use was not to last. IPTO's original mandate included the goal of eventually connecting its funded sites, even prior to the specific proposal of the ARPANET~\cite{leiner1997pastandfuture, robertswessler1970resourcesharing, marrill1966network}.\footnote{The ARPANET proposal ultimately covered more specific goals: a network for load sharing, a message service, a remote service, data and  %sharing, 
program sharing, specialized systems software and hardware,   %specialized systems software, 
and scientific communication. It also  %and 
described the %basic 
operation of such a network to ``foster the `community' use of computers''~\cite[p. 2]{roberts1967arpanetproposal}.} By the end of 1971, when ARPA was completing the first phase of the ARPANET's construction to connect 15 Interface Message Processors (IMPs), IPTO's vision became a funding precondition~\cite{heart1970imp}. At this time, IPTO appealed to its contractors to no longer just use their computers as local lab resources, but rather to exercise their connection to the network~\cite[pp. 44-46, p. 55]{abbate1999inventing}~\cite{robertswessler1970resourcesharing}.\footnote{This notion of resource-sharing was, at least at first, considered the distributed analogue of the current local time-sharing computing paradigm, which had its own accounting concerns:  ``The  goal  of  the  computer  network  is for  each  computer  to  make  every  local  resource  available  to  any  computer in the net in such a way that any program available  to  local  users  can  be  used  remotely  without degradation. That  is,  any program  should  be  able  to  call  on  the resources  of  other computers much  as it would  call  a  subroutine.  The  resources  which  can  be  shared  in  this  way  include  software and  data,  as  well as  hardware.  Within  a  local  community, time-sharing systems already permit the sharing of software resources. An  effective  network  would  eliminate  the  size and distance limitations on such communities''~\cite[p. 543]{robertswessler1970resourcesharing}. See also~\citet[p. 589]{carr1970rfc33pub}:  %(a conference version of ~\citet{rfc33}:
``However, early time-sharing studies at the University of California at Berkeley, MIT, Lincoln Laboratory, and System Development Corporation (all ARPA sponsored) have had considerable influence on the design of the network. In some sense, the ARPA network of time-shared computers is a natural extension of earlier time-sharing concepts.'' Also, see generally~\citet{marrill1966network}.} All 15 of these sites housed other ARPA-funded computing projects. As a result, even if networking was not the focus of every site's individual projects, each site was expected to participate~\cite[p. 50, pp. 77-78, p. 161]{abbate1999inventing}~\cite[p. 548]{robertswessler1970resourcesharing}. In other words, IPTO needed its contractors to utilize the network they had invested in building, in order to test the network's potential for distributed computing—for two or more remote nodes to effectively work together to complete computational tasks. So, starting in late 1970, the perception of ``no strings attached'' funding began to crumble~\cite{rfc75,rfc77,rfc82,rfc101,rfc231}. It was becoming clear that ARPA's funding did in fact come with a particular yoke: Contractors did not just have to connect to the ARPANET; they also had to use it.

This pivotal moment, nearly 10 years after IPTO's founding, marked when it was possible to move theory to practice---to empirically validate IPTO's commitment to resource-sharing via the early ARPANET. Historian Janet Abbate discusses how this promise faded rather quickly: ``the decline of the ideal of resource sharing,'' she argues, came as the result of ARPANET's usability issues~\cite[p. 104]{abbate1999inventing}~\cite[p. 5]{rfc369}. While connecting to the network had been a grueling engineering task, it ultimately was just the beginning of ARPANET's resource-sharing challenges. Once connected, it was difficult to locate specific resources in the network and lingering interoperability issues meant that, even once a resource was found, it often remained unclear how to access it~\cite[pp. 5-6]{rfc82}~\cite[p. 6]{rfc369}~\cite{rfc531}. Abbate thus concludes that resource-sharing seemed more onerous than it was worth, such that the ``demand for remote resources fell,'' leaving ``many sites rich in computing resources...looking for users'' and the ARPANET a technology in search of an appropriate application~\cite[p. 104]{abbate1999inventing}.  

But the usability of a network is itself constructed through choices regarding its administrative infrastructure. Seeking to unpack Abbate's invocation of ``usability,'' we argue that the challenges of resource-sharing, and the changes it produced in IPTO-funded computers' use, can also be ascribed to seemingly mundane (but in fact very difficult) issues of bookkeeping. Even though ARPA continued to foot the entire bill for the ARPANET, both in terms of capital and communication costs~\cite[p. 85, p. 161]{abbate1999inventing}, from the perspective of individual research sites, resource-sharing constituted a sacrifice---a loss of the unrestricted, free local computer use that had been the status quo.\footnote{It is possible that this was just a \emph{perceived} sacrifice of local resources for collective external use, with no actual scarcity of computing resources for those that wanted it. Nevertheless, even if not an actual sacrifice, as we discuss, there was a reluctance to even agree to share local resources with the network.} That is, by requiring sites to reallocate a portion of their computing resources for distributed use, resource-sharing seemed akin to ceding control of one's own local computing budget to remote users with their own respective, perhaps even competing, needs~\cite[p. 50]{abbate1999inventing}.

As a result, even though IPTO required them to use the ARPANET, many contractors exhibited unwillingness to do so, wondering how to prioritize local and remote use. Richard G. Mills, the director of MIT's information processing services, succinctly captured this hesitancy, saying: ``There is some question as to who should be served first, an unknown user or our local researchers''~\cite[p. 226]{abbate1999inventing}. ARPA was no longer covering all operational costs: It was not paying for the coffee and pastries, and it was not compensating for the loss of previously unrestricted local resources now being shared with others in the network. Inducing resource-sharing therefore exposed a fundamental, underlying tension in distributed computing. Individual labs may not have wanted to give up their local resources, but they also recognized the potential value of being able to use other labs' resources. One suggested solution to this tension was for site administrators to bill remote users in order to recoup losses or disincentivize remote use~\cite[p. 7]{rfc82}. As J. Pickens asserted in RFC 369, ``if distributed computing is allowed, then distributed billing is a necessity''~\cite[p. 6]{rfc369}.

For labs to charge for the use of their resources represented a fundamental shift in the management paradigm of the ARPANET. Until this point, there was no need to do low-level accounting of individual line items of resource usage---of who was using (or even misusing) specific parts of the network---because ARPA was managing the overarching cost. Billing, however, now imposed a new burden of operational costs on individual sites, in which low-level bookkeeping was going to become crucial for the first time. In the remainder of this section, we show how the challenges presented by the need to develop accounting mechanisms structured the ambivalence towards resource-sharing in the early ARPANET. Accounting posed a non-trivial problem. Local time-sharing computers already ``possess[ed] elaborate and definite accounting and resource allocation mechanisms''~\cite[p. 5]{rfc33}, but these did not naturally extend to the distributed resource-sharing environment.\footnote{As RFC 504 asked, ``If you employ accounting Procedures that require cost recovery, how, if at all, should they be modified to work in a network resource sharing environment?''~\cite[p. 4]{rfc504}. This was not at all straightforward for key technical reasons because inherent issues with consistency between nodes in a distributed system further complicate the mechanics of correct accounting. While consistency in distributed databases is now known to be a topic of fundamental importance in distributing computing, it seems that this issue was first noted in RFC 677~\cite{rfc677} in relation to maintaining duplicate databases for correct, consistent Terminal Interface Processor (TIP) accounting. RFC 677 notes that its contents go beyond ``ARPA-like networks'' and ``are generally applicable to distributed database problems''~\cite[p. 1]{rfc677}, talking about issues of partition tolerance~\cite[p. 3]{rfc677} and consistency~\cite[p. 4]{rfc677}, as well as how timestamps can be useful for maintaining consistency because they are monotonically increasing (noting, however, that this can be complicated by clock skew between nodes). For a more contemporary treatment of these issues, refer to~\citet{abadi2012tradeoff}.} The non-triviality of distributed accounting posed challenges that researchers neither knew how to solve, nor really desired to spend time solving in place of performing their individual research.

\subsection{Explicit and implicit billing: The case of ``free'' file transfer}\label{sec:freeftp}

The early ARPANET debate over ``free'' file transfer gives an intuition for the complexity of accounting in the distributed resource-sharing environment. File transfer was (and remains) one of the basic functions of resource-sharing over networked computers. It facilitates the transfer of files from one node to another, enabling %distributed 
sharing among remote users. The File Transfer Protocol~\cite{rfc354} first described this capability, which requires resource usage, including memory and CPU utilization, and thus made it a prime candidate feature for billing.\footnote{To enable billing, RFC 385 added an account (ACCT) command to the FTP protocol, in order to distinguish accounts used for resource accounting as serving a different function from users logged onto the network~\cite{rfc385}.} Nevertheless, despite the apparent necessity for billing to support the network, many individual users wanted to avoid payment. When using FTP, they leveraged a loophole: the MAIL FILE feature, which allowed for bypassing login on the remote host that housed the file of interest, and made it possible for a user to mail the file to themselves, in place of transferring it via TELNET-based connections.

In response to the popular use of FTP's accounting loophole, Rob Bressler, a network protocol developer at BBN, issued an RFC calling to codify a more appropriate accounting-free FTP use pattern: the implementation of free, loginless file transfer, which would give users the free access they wanted without abusing FTP's intended use~\cite{rfc487}. Rather than using MAIL FILE, his proposal expressly allowed for users to bypass authentication via a  %deliberate 
loginless facility and, without logging in, it would not be possible to account for who was transferring the file. The transfer would be ``free,'' as it would not be possible to bill an account for it.\footnote{On hardware for which such loginless access was not possible, such as TENEX machines, Bressler proposed adding an account named ``FREE,'' which could be used by users to avoid billing to specific accounts~\cite{rfc487}.} Bressler's proposal for an intentional, free file transfer feature immediately raised questions about what it meant for resource usage to be ``free''~\cite{rfc491}. Notably, fellow BBN network engineer, Ken Pogran rebuffed Bressler's RFC for making ``sweeping assumptions...about the nature and use of accounting mechanisms~\cite[p. 1]{rfc501}.'' Pogran resolved to ``un-muddle'' so-called 'free file transfer','' making the case that the resource usage involved in FTP (deemed by Bressler as negligible) was in fact quite costly. While CPU utilization for transferring one file might seem negligible, over time such costs add up and certainly cannot be called ``free''~\cite[p. 4]{rfc501}. In short, Pogran argued, nothing is free if you are the one who has to worry about costs.

Yet, while Pogran challenged what it meant for resource usage to be called ``free,'' he did not claim that such actually not-``free'' usage should be disallowed. Rather, he contended that ``free'' should mean that resource usage was free of charge for a user at a research site---and that such ``free'' usage should get charged to an overhead ``network services account'' that ultimately got billed to ARPA~\cite[p. 3]{rfc501}. In other words, while Pogran seemed to have taken a more nuanced view than Bressler about the costs of resource-sharing, RFC 501 does not ultimately ``un-muddle'' the issue of distributed accounting. Like Bressler, Pogran also saw merit in %the approach of 
avoiding the particulars of accounting; he, too, found it desirable for researchers like himself to not be concerned with the minutiae of how costs got covered. However, unlike Bressler, Pogran made explicit that ignoring costs did not simply make them go away. Rather, he highlighted how explicitly implementing mechanisms to evade accounting corresponded to the status quo of ARPA being on the hook for the bill. The exchange between Bressler and Pogran underscores the same contradiction: Both talked about accounting as necessary to recoup local site losses due to resource-sharing, but both also affirmed the common desire of ARPANET researchers to not pay to use remote resources. The \emph{ad hoc} strategy of ``free'' accounts would prove infeasible in the longer term, when it was expected for the network to host nodes and users not funded by ARPA~\cite[p. 3]{rfc501}.

\section{Accounting as measurement: Contesting the necessary functions of networked computing} \label{sec:measurement}

The debate over ``free'' file transfer demonstrates that deciding how to classify what resource usage needed to be accounted for was a challenging and contentious problem.\footnote{See Appendix~\ref{app:sec:researchservice} for discussion of similar attempts at this classification.}
As IPTO pushed for resource sharing more actively, it was not clear what needed to be accounted for. Making the decision to explicitly build workarounds to avoid accounting---%folding
pushing some types of resource-sharing costs back to ARPA---ultimately obscured the complex and pervasive role of accounting in a distributed, resource-sharing environment.

The architects of the ARPANET repeatedly punted on designing mechanisms for accounting. Even during  the ARPANET's earliest years, around the completion of the connection of the first part of the network in 1971, accounting had frequently come up as a necessary function, albeit one with unclear requirements~\cite{rfc75, rfc77, rfc82, rfc101}. In response, Bob Kahn, co-lead of the ARPANET protocols team, prepared RFC 136: the first unified attempt to clarify accounting's role in resource-sharing.\footnote{RFC 136 was concurrent with discussions about how to distinguish between billed Research Centers and free (but limited) Service Centers (Appendix~\ref{app:sec:researchservice}). Notably, this predates the debate over ``free'' file transfer discussed in Section~\ref{sec:freeftp}, indicating that the problems RFC 136 raised remained unresolved and carried over into later debates such as FTP.} He raised ten, as-yet-unanswered crucial questions related to accounting, which ranged from a future of potential private control of the ARPANET through government regulations of network use and to how resource usage should be measured and characterized. These questions, in attempting to clarify what accounting requires, instead serve to clarify just how complicated and expansive accounting is: While accounting clearly involves billing, billing is not the only component of accounting.

Several of Kahn's RFC 136 questions concerning accounting implicated the ability to take measurements of network activity. While questions and engineering activities concerning measurement predate RFC 136~\cite{rfc100}~\cite[Appendix]{mckelvey2018internet}, this was the first time that an RFC unified questions about measurement and accounting within the same scope.\footnote{``The method of network operation and the potential for its growth are relevant factors to be considered in formulating a plan for Host accounting.  For example, the answers to the following questions provide a useful background for reference: 1. Who or what operates the Network? 2. What is the criteria upon which new sites should be incorporated into the Network? 3. What regulations, if any, apply to the connection of non-ARPA sites? 4. What is the relation, if any, between the ARPA Network and common carrier services? 5. What procedures are required to bring new sites on board and up to speed? 6. What is the most effective way to characterize their Resources? 7. What usage of other Network resources do they anticipate? 8. What procedures will be required for a typical user to obtain access to that Host? 9. What is their charging policy and for what items? 10. Are their rates in accordance with government standards?'' ~\cite[p. 1]{rfc136}. In short, adding an ``account'' field to different network functions perhaps could help facilitate accounting, but it 
was not sufficient on its own to capture the wide-ranging semantics of accounting implicated by Kahn's considerations.} %, outlined in RFC 136.} 
Prior to this RFC, accounting and measurement had generally been considered separate---albeit both necessary---functional concerns~\cite[e.g., p. 19]{rfc100}~\cite[Appendix]{mckelvey2018internet}, neither of which had been solved by the ARPANET architects. While accounting was conceived of as billing and treated as a nuisance to be kept separate from research (Section~\ref{sec:freeftp}, Appendix~\ref{app:sec:researchservice}), measurement was afforded the status of being integral to research. Measurement was a necessity for those ``interested in the network as an object of study,''~\cite[p. 2]{rfc77}~\cite{rfc82} while accounting was not considered to have such a central role.

The earliest example of measurement's importance as a research function concerns the work of Gerry Cole at UCLA. As late as February 1971, Cole collected and analyzed network data in order to better understand resource usage patterns in the ARPANET's novel distributed environment~\cite[p. 2]{rfc101}.\footnote{One %early
RFC noting this project read: ``Gerry requested that when people are set up to use the Network, they inform him so that he can gather statistics. UCLA will eventually have a program to scan the Network for utilization, but if people could tell him when they were going on to use the Network, it would be easier to measure meaningful things and interpret the data from a knowledge of type of usage''~\cite[p. 2]{rfc101}.} Shortly after these initial measurement efforts, BBN took over the role more formally and set up the Network Control Center (NCC), led by Alex McKenzie, to measure network statistics. The NCC monitored all nodes attached to the ARPANET at this time, and assumed the role of ensuring that the entire network ran smoothly by documenting reliability issues, debugging and diagnosing malfunctions, and monitoring resource usage on the network~\cite[pp. 64-67,72]{abbate1999inventing}.\footnote{In RFC 101, %Bob 
Kahn was recorded to have mentioned that BBN had an interest in collecting measurements on the network~\cite[p. 2]{rfc101}. Abbate %'s work 
confirms this, %. She documents 
documenting Alex McKenzie's role at BBN: He joined the ARPANET project when BBN's  %ARPANET 
node went online as part of the November 1970 deployment %push
~\cite{rfc77, rfc82}, and in 1971 he took charge of the NCC~\cite[pp. 64-67]{abbate1999inventing}:  ``the NCC acquired a full-time staff and began coordinating upgrades of IMP hardware and software. The NCC assumed responsibility for fixing all operational problems in the network, whether or not BBN's equipment was at fault. Its staff monitored the ARPANET constantly, recording when each IMP, line, or host went up or down and taking trouble reports from users. When NCC monitors detected a disruption of service, they used the IMP's diagnostic features to identify its cause. Malfunctions in remote IMPs could often be fixed from the NCC via the network, using the control functions that BBN had built into the network''~\cite[p. 65]{abbate1999inventing}; ``By 1976, the Network Control Center was, according to McKenzie, 'the only accessible, responsible, continuously staffed organization in existence which was generally concerned with network performance as perceived by the user.' ... The NCC had become a managerial reinforcement of ARPA's layering scheme''~\cite[p. 66, internal citations omitted]{abbate1999inventing}.} The NCC was not limited solely to conducting research on network utilization, but was responsible for its \emph{de facto} operation, a role we discuss in greater detail in Section \ref{sec:management}.

As Bob Kahn suggested in RFC 136, these functions, aside from playing a role in developing an understanding of network use, were an essential component of accounting~\cite[p. 1]{rfc136}.  He made clear that it would not be possible to account for resource usage without having appropriate mechanisms in place to measure resource usage. Accounting-related measurements would not just involve the ``who, what, where, and when'' of network use~\cite[p. 3]{rfc585}; they would also involve metrics concerning site performance, such as user response times and frequency of crashes~\cite{rfc369}.\footnote{%Such
Metrics would be important for accurately accounting for past resource usage~\cite[p. 5]{rfc585}, and would come to be recognized in the 1980s as important for enabling individual sites to predict future resource usage and corresponding costs~\cite[pp. 1-3,8,27,44]{rfc869}.} By indicating overlap with functions like measurement, the accounting-related considerations Kahn raised in RFC 136 implicated fundamental questions about what the network should do, and how it should be implemented.

These considerations raised not only practical challenges, but also ideological questions about the network's purpose. The act of performing accounting itself consumes network resources; it costs something to account for costs. The early ARPANET architects considered these costs to be overhead. They were concerned that accounting ``costs space''~\cite[p. 3]{rfc82} and could cause ``undue delays in accessing distributed resources''~\cite[p. 4]{rfc592}. They viewed the resource consumption involved in accounting as an imposition that was in tension with the ARPANET's fundamental, ideal goal to achieve distributed resource sharing~\cite[pp. 96-97]{abbate1999inventing}~\cite{robertswessler1970resourcesharing}. Accounting therefore presented a seemingly irreconcilable contradiction: On the one hand, ARPANET architects repeatedly acknowledged that \emph{accounting was critical} to facilitate interconnection between distributed nodes; on the other, they viewed \emph{accounting as hostile} to that very same goal.

This view of accounting as a burden to the network can be understood in relation to the End-to-End architectural principle guiding the construction of the ARPANET. As mentioned in Section~\ref{sec:intro}, from a technical perspective, End-to-End, is a preference toward parsimony in the network---of placing application-specific functionality where it is needed at the end hosts, rather than implementing it inside the network as a feature accessible to all hosts. This principle biases toward only placing the essentials for connectivity inside the network, with the notable exception for features ``justified only as performance enhancements''~\cite[pp. 1,9]{saltzer1984endtoend}. Accounting certainly could not be considered a performance enhancement.\footnote{``The principle, called the End-to-End argument, suggests that functions placed at low levels of a system may be redundant or of little value when compared with the cost of providing them at that low level. ... Low level mechanisms to support these functions are \emph{justified only as performance enhancements}''~\cite[p. 1, emphasis added]{saltzer1984endtoend}. The ability to record measurements, which we note is fundamentally tied to accounting, is essential from the ideological perspective of End-to-End; it is not possible to justify performance enhancements without being able to measure performance.} In fact, as we have seen, ARPANET architects viewed accounting as a performance hit that should be kept ``to a minimum''~\cite[p . 3]{rfc82}. Abiding by End-to-End, it would be ``uneconomical'' to include accounting within the network, rather than pushing its implementation to end hosts. The reluctance to implement accounting was therefore not just at the level of specific engineers who wanted to evade paying for FTP; rather, it reflected the significant ideological challenges that accounting presented for a network priding itself on its parsimony.

Consistent with an End-to-End approach to accounting, %Kahn's 
RFC 136 was scoped to ``Host Accounting.'' Its intent was to address the issue as a matter for the end nodes rather than inside the network. Yet, toward the end, %of the RFC, 
Kahn posed a speculative question for the future, which can be read %as 
in direct tension with End-to-End: ``Should Host accounting information \emph{eventually flow via the Network}?''~\cite[p. 4, emphasis added]{rfc136}. However, since accounting was clearly not a ``performance enhancement''~\cite[p. 1]{saltzer1984endtoend}, for accounting to not violate End-to-End, it would have to be justified as an essential architectural feature of connecting distributed nodes. An attempt at such a justification would have run contrary to the RFCs at this time, which always list accounting as distinct from the technical requirements of connectivity~\cite{rfc75,rfc77,rfc82,rfc101}.\footnote{Given that the architects commonly recognized the importance of accounting---%as was clear both 
seen in Kahn's own RFC 136 and those that preceded it~\cite{rfc75,rfc77,rfc82,rfc101}---it was arguably reasonable for Kahn to question whether accounting was fundamental enough to be an in-network feature. %In this context, 
One could conceivably read Kahn's question as a provocation: Will there be a time at which accounting will be important enough to violate End-to-End? This would have been unthinkable %to consider viable 
in 1971. Instead, Kahn recommended a more gradual, conservative approach: ``the implementation of standard automated accounting procedures involving the use of the Network will be deferred until non-automated procedures have been understood and stabilized. Early experimentation in this area is appropriate, however''~\cite[p. 2]{rfc136}. Even after RFC 136, accounting continued to be punted until some unspecified, but inevitable point in the future. RFC 82 acknowledges that one possibility would be to ``worry about accounting when saturation occurs''~\cite[p. 7]{rfc82}---i.e., when resource usage reached an extent for which it would be absolutely necessary to do accounting, especially since at that time ``non-ARPA folks [would] be able to connect''~\cite[p. 4]{rfc136}. Accounting could also be put off until a future in which ARPA was no longer responsible for the network infrastructure---when another government agency or a private commercial entity assumed the ``total cost of operating''~\cite[p. 2]{rfc136}. It is thus reasonable to conclude that, even as early as 1971, commercialization loomed as a possible future for the ARPANET, at which point accounting would be absolutely necessary for reimbursement ``on both a connect and usage basis''~\cite[p. 4]{rfc136}.} In the interim, there remained no concrete plan %in place 
for developing accounting procedures.  

\subsection{The policy of no policy takes hold}\label{sec:nopolicy}

In the years %that followed 
following RFC 136, the ARPANET fell short of achieving IPTO's goal of facilitating resource-sharing~\cite{roberts1967arpanetproposal}. Abbate's canonical narrative attributes the decline of this ideal to the fact that the ARPANET was very difficult to use, especially for new users trying to join the network. She argues that this created an identity crisis for the ARPANET; it was a technology in search of an application, a role that email was well-posed to assume as it did not suffer from the same usability issues that plagued resource-sharing~\cite[p. 106, the ``smash hit'' of email]{abbate1999inventing} (%See 
Appendix~\ref{app:sec:abbateusability}). While usability issues presented a challenge for the adoption of resource-sharing, we argue that the decline of resource-sharing can also find its roots in the inability to account for resource usage. With no mechanism to account for resource usage, it was not possible to realize the imagined potential of resource-sharing. To reiterate and rephrase Pickens' assertion, without the necessary functionality of distributed billing, distributed computing was not feasible~\cite[p. 6]{rfc369}.

Accounting was fundamental to the network, but was also fundamentally unresolved. It was non-trivial in all the same ways that devising protocols for the use of shared resources was non-trivial: inescapably tied to measuring and managing network performance, and implicated in everything from resource allocation to quality of service. These tensions %surrounding accounting 
extended well beyond the early 1970s~\cite[e.g.]{rfc505, rfc677, rfc759, rfc892, rfc1017}. In 1994, one RFC called accounting (grouped with security) ``the bane of the network operator,'' but admitted that it is the feature ``most requested...by those who are responsible for paying the bills''~\cite[pp. 149-150]{rfc1716}. More recently, David Clark referred throughout his latest book to the importance of accounting, and yet also relegated it to a secondary function, saying it ``only plays a supporting role compared to the core objective of forwarding data, and researchers like to work on the lead problem, not the supporting role''~\cite[pp. 46-47]{clark2018internet}. Importantly, as we argue in the next sections, these accounting tensions carried over to network-wide policy tensions. There were repeated attempts to minimize or redefine the role of accounting, often in the service of deprioritizing its implementation. In the process of attempting to preserve the ``free'' aspects of the early ARPANET, the network became a free-for-all.

\vspace{-.09cm}
\section{Accounting as management: Tracing the boundaries of responsibility and authority} \label{sec:management}

The ambivalence toward accounting's role in the network persisted in the decade that followed the ARPANET's expansion to more sites. Our discussion of the early years of the network argues that, even during a time of relatively low saturation, accounting emerged as a function with bearing on both billing and measurement. But by the mid-1970s, users' demands for a more reliable network to support their research work showed how without management the network was not sufficiently operational. Later, the ARPANET's connection with other networks that constituted the Internet, the ``network of networks,''\footnote{%During 
In the 1980s, ``network of networks'' was used interchangeably with %the term 
``Internet'' to emphasize the transition from talking about the ARPANET specifically to the workings of connecting ARPA's sites with other existing networks. %We discuss this shift later in this section. 
For example, in sketching out a proposal for an interagency research institute, Barry Leiner referred to the ``'Network of networks' or Internet model of interconnection"~\cite[p. 3]{rfc1015}.} put the possibility of interacting with mistrusted agents front of mind for the network protocol developers. This section explores how accounting, the flexible term for many administrative aspects of network development, was initially contrasted with the core need to create features for network \emph{management}. By the late 1980s, however, descriptions of what would constitute effective network management explicitly referred to accounting as a main component. If the early case of FTP shows how even adding a data field that would allow for future billing carries with it meaningful decisions about the purpose of the network, the 1980s Internet discussions over network management proved accounting raised serious questions about the boundaries of responsibility for and authority over elements of a networked system.

\subsection{Management as failure: The visibility of humans at the Network Control Center}

As Section~\ref{sec:measurement} explored, there was no obvious way to measure activity on the network and implementing accounting required tackling various trade-offs (for example, losing data). But over time, ARPANET engineers also faced changing expectations of users from the network, dictating new needs for accounting that supported diagnostics. Alex McKenzie, who led the measurement efforts at BBN and established the Network Control Center (NCC) there (Section~\ref{sec:measurement}), later reflected on the ARPANET's growing pains as users came to expect greater operational reliability from it: ``Once a set of host protocols were defined to allow connections…it was remarkable how quickly all of the sites really began to want to view the network as a utility rather than as a research project''~\cite[p. 11]{anderson1990mckenzie}. Meaning that, although ARPANET was itself envisioned as an ongoing research project in network communication, it also became the infrastructure that supported other research activities. Allowing that level of reliability required proactive management of use issues.

Between 1972 and 1976, the NCC became the focal unit for troubleshooting issues on the ARPANET, gradually expanding support to additional sites %the coverage of sites for which they offered support, 
alongside the network's expansion~\cite[p. 15]{anderson1990mckenzie}.\footnote{%Alex 
McKenzie later speculated that he was tasked with running the NCC during this period because of his personal belief that the ARPANET should become more of a utility focused on operational issues, rather than an experimental research project that tolerated routine disruptions---to ``take it very seriously if anything was broken.'' For context, the extent to which ``things were broken'' can be seen in some data %put forward 
in RFC 369, ``Evaluation of ARPANET Services,'' which surveyed a couple of months in early 1972, just preceding McKenzie's role at the NCC. The survey highlighted that the reported mean time between failure was at best 2 hours and at worst 5 minutes. The average of time of ``trouble free operation'' amounted only to 35\%, which the RFC described as ``a figure untenable for regular user usage''~\cite[p. 4]{rfc369} Later on, McKenzie reported that the NCC considered their efforts to treat the network as a utility successful when providing reliability 98-99\% of time, which he conceded was not comparable for what constituted reliability for a utility such as electricity~\cite[p. 15]{anderson1990mckenzie}.} It stepped into this role ``because the users really needed a single point of contact'' in case any component of the network failed.\footnote{We previously discussed Abbate's argument about the ``decline of the ideal of resource sharing'' as a matter of neglected usability issues (also see Appendix~\ref{app:sec:abbateusability}). The role that the NCC occupied during this time highlights our point about the ways accounting activities (in this case, measurement data that informed diagnostics) constitute a significant aspect of ``usability.'' But it is also worth noting that the NCC's responses to users' demands did not concern creating interfaces for finding resources or navigating their use, but, more explicitly, issues users had with figuring out whether a specific device or service was down for some reason. McKenzie describes the following scenario: ``It's absolutely no use to some geophysicist in Washington who's trying to do something in the DARPA seismic monitoring program, to say, `Well, you can call the Network Control Center, if that doesn't work call MITRE and ask about their modems, and if that doesn't work call ISI, and ask about their computer.' Nobody would work that way, I wouldn't work that way myself''~\cite[p. 13]{anderson1990mckenzie}. Even when we focus our attention on the experience of users of the network, rather than its engineers, we see that accounting plays a significant role in the ongoing ability to access services reliably.} The bind facing the NCC team, as McKenzie put it,
was that ``even though [their] authority didn't expand, [their] responsibility expanded quite a bit''~\cite[p. 13]{anderson1990mckenzie}. While the NCC had no authority over how a host on the network in another institution operated, they still found themselves answering for issues that arose with its use. The lack of authority that characterized the NCC's efforts to ``control'' the network pointed to the need to develop clearer protocols for network management.

While the NCC was committed to addressing network failure through proactive management, some ARPANET engineers believed that management itself was a sign that the research project of ARPANET was a failure. In his 2018 book, David Clark, devoting an entire chapter to the topic, defines management as ``those aspects of network operation that involve a person''~\cite[p. 260]{clark2018internet}. Defined in this way, through the need for human intervention, management is already set up to contrast with what network engineers might consider the core of their work. Clark, who clearly has come to view management as an important aspect of networking, points out that some engineers see management as a signal for design failure (as opposed to simply the temporary failure of a component of the network) because ``a properly designed network should run itself''~\cite[p. 260]{clark2018internet}. In other words, developing the infrastructure for accounting that would aid the NCC's management role was to admit that the ARPANET required such ongoing, labor-intensive human operation.\footnote{By the mid-1970s, the NCC was a 3-person team during East and West Coast business hours, and was further operated by 1-2 people at other times, to offer 24-hour, 7-day-a-week coverage~\cite[p. 13]{anderson1990mckenzie}.}

From his present-day vantage point, Clark describes the network engineers who consider management as a necessary aspect of any robust network as ``pragmatists.'' Throughout this paper, our argument has been that designing accounting mechanisms that would enable accountability is especially key for distributed technical systems: In the case of network failures, accounting would allow for inspection after-the-fact. Recognizing human operators performing ``management'' as part of the network is, therefore, not giving up on the project of network computing, but acknowledging the range of design decisions that have to be addressed to develop appropriate infrastructure.\footnote{%For example, 
Already in 1981, Jon Postel proposed the Internet Control Message Protocol (ICMP), a means to send back information about a lost datagram, recognizing that the network's protocols are ``not designed to be absolutely reliable''~\cite[p. 1]{rfc777}. %These accounting functions of 
ICMP's accounting functions allowed individual users to probe the network to identify when a particular host was down, and is used to this day as a ``ping'' by both professional network managers and ordinary Internet users% of the 
~\cite[p. 263]{clark2018internet}. But such minimal diagnostics implementations still provided a rather limited amount of management tools.} Throughout the 1980s, developing protocols to support recording data relevant to management became increasingly more mainstream at the IETF. In 1988, Vint Cerf circulated a memo on behalf of the Internet Activities Board recommending the adoption of a common network management framework to avoid incompatibility issues across the network~\cite[p. 4]{rfc1052}. This marked a decade-long shift from implementing a patchwork of accounting and monitoring tools to developing an entire conceptual language around the issues of network management.

The new framework divided the meaning of management into five functional areas: ``fault management, configuration management, performance management, accounting management, and security management,'' later formalized as the FCAPS framework~\cite[p. 8]{rfc1095}~\cite[p. 261]{clark2018internet}.\footnote{For details on the ``Standards War'' that took place concerning this management framework, refer to \citet{russell2006war}.} The framework at once narrowed the role of accounting management, defining it as making ``it possible to charge users for network resources used and to limit the use of those resources,''~\cite[p. 8]{rfc1095} and expanded its centrality by implicitly recognizing that the tools needed to allow for accounting management were also needed to support a variety of other management functions.

Even in the seemingly discounted role of ``accounting management'' within the FCAPS framework, we can already see hints to the types of management decisions that went beyond network failure diagnostics. We %have 
previously argued that accounting, conceived as billing, already surfaced a variety of questions about what it means to share resources over a network. Accounting as management recognized the potential need to limit access to resources. The developing language around network management in the 1980s marked a transition from the early days of management, consisting of identifying failing components on the network and troubleshooting them, to allowing %network 
managers to make decisions about network traffic and resource access, to prioritize some users over others.

\subsection{The ``network of networks'': Management without trust}\label{sec:management:notrust}

As the scale of the network grew, not only adding new sites to the ARPANET but also incorporating connections with other existing networks, differences in management across these domains came to the fore.\footnote{Throughout this paper, we focus on the ARPANET as a precursor to the Internet, and the differences between the two, while both technical and administrative, are primarily outside the scope of our discussion of accounting and accountability in the development of network computing. However, during the 1980s, the ARPANET,  thanks in part to management efforts such as the NCC, was considered relatively reliable; it was internetworked with other, less reliable networks to constitute the early version of what we may consider as ``the Internet.'' For more on the architectural and institutional challenges involved in creating the Internet, see ``Chapter 4: From ARPANET to Internet'' in~\citet{abbate1999inventing} and ``Chapter 7: Alternative Network Architectures'' in~\citet{clark2018internet}.)} Throughout this transition to a ``network of networks,'' the applicability of a particular management design captured in the FCAPS framework came to define the boundaries of ``Administrative Domains'' (ADs)---sectors of a network defined precisely through their shared management~\cite[p. 8]{rfc1095}.\footnote{While calling different units of the network ``Administrative Domains'' (ADs) and at times ``Administrative Regions'' (ARs) emphasized that conflict or incompatibility on the network often stemmed from the differences in management priorities and systems, over time the term was replaced with ``Autonomous Systems'' (ASs), a term that obfuscated the type of management considerations we explore in this section.}  In our discussion of the developing language around management and accounting above, we have treated the kinds of traffic and resource access issues that occur within an AD. But the introduction of \emph{inter}-AD management raised a new set of concerns and competing interests which needed to be accounted for.

In 1983, Barry Leiner took over directing the Internet research program at ARPA and led a broader effort of organizational restructuring, which both he and Clark have attributed to the growth of the network~\cite[p. 50]{russell2006war}~\cite[p. 29]{leiner2009briefhistory}.\footnote{In Leiner's 2013 posthumous induction into the ``Internet Hall of Fame,'' the citation credited him as having ``helped set up the bureaucratic structures that developed Internet communication protocols''~\cite{halloffame2013leiner}. These bureaucratic structures included the establishment of the Internet Activities Board (IAB) and the task-force structure for tackling specific technical aspects of the network's development, which ultimately  resulted in the IETF. Historian Andrew Russell argues that the Internet's technical standards governing structure was itself a political innovation, often overlooked by narratives that focus on the innovative technical aspects of the Internet~\cite{russell2006war}. It is therefore not surprising for Leiner to emerge as one of the more prolific RFC authors that focused on the administrative and accounting needs of the growing network.} In a series of RFCs during his tenure, Leiner became a vocal commentator on a variety of accounting concerns that the new inter-AD age raised. In an ``idea paper'' circulated as an RFC in 1987, Leiner used the experience of the growing Internet to sketch out the challenges facing any kind of interagency research network. The conceptual problem at the core of this type of network remained the lack of a ``consistent mechanism to allow sharing of the networking resources''~\cite[p. 1]{rfc1015}. In framing the issues facing a ``network of networks'' in terms of the needs of different research agencies running their own networks, or ADs, Leiner paid close attention to the possibility that %these
intra-AD priorities might conflict with the inter-AD design: ``[to] assure appropriate accountability for the network operation, the mechanism for interconnection must not prevent agencies from retaining control over their individual network''~\cite[p. 9]{rfc1015}. In 1972, Alex McKenzie found himself responsible for the performance of individual components of the network without the authority to manage them. Here, over a decade later, Leiner raised the reverse concern---that individual agencies would not be able to be accountable to what occurred on their networks due to a lack of authority over aspects of internetworking activity. Leiner called for a ``management approach'' that would allow local control while still sharing resources with users sponsored by other agencies~\cite[p. 11]{rfc1095}. In Leiner's description, necessary tools for individual network managers included both user access control and privacy, and accounting mechanisms ``to support both cost allocation and cost auditing''~\cite[p. 9]{rfc1015}. The shift to a ``network of networks,'' then, only heightened what has already been a persistent, unresolved need for accounting as part of a network management approach.

At the interface between the different ADs of the "network of networks" sit special host computers with the responsibility to route and ability to translate data across domains, called gateways~\cite[p. 6]{rfc1015}~\cite[pp. 128-129]{abbate1999inventing}. Gateways acted as ``buffers'' between ADs, simplifying the need for each AD to have working knowledge of others~\cite[p. 129]{abbate1999inventing}. The Exterior Gateway Protocol (EGP), introduced in 1982, was a means of allowing this patchwork of ADs to appear to an end-user as ``a single internet''~\cite[p. 3]{rfc827}. EGP contained mechanisms for discovering ``neighboring'' ADs and allowed those ``neighbors'' to exchange reachability information---to know whether the ``neighbor'' was open to receiving traffic. A key feature of  %the 
EGP was to enable ``each gateway to control the rate at which it sends and receives network reachability information, allowing each system to control its own overhead''~\cite[p. 5]{rfc827}. The resource-sharing concerns we first encountered in Section~\ref{sec:billing}, now cast against the changing Internet environment that expected increasing numbers of internetworked ADs, took on a new aspect of accounting as management---a baked-in assumption of mistrust between the different networks. If one could not account for users of other networks, then a buffer between these ADs would be needed.

A few years later, David Clark described the purpose of %the 
EGP as follows: ``to permit regions of the Internet to communicate reachability information, even though they did not totally share trust''~\cite[p. 2]{rfc1102}. This assumption of mistrust cut in two directions: On the one hand, an AD might wish to limit the amount of information other actors on the network would have access to, using gateways as a buffer; on the other, mistrust also required a higher degree of accounting information to allow for individual AD management decisions. EGP was an imperfect solution. It limited the amount of information different ADs had to share in order to still make use of the ``network of networks,'' but left under-developed the tools of network management and accounting that would allow for the kind of responsibility Leiner was advocating for.\footnote{In 1989,  %a new protocol, 
the Border Gateway Protocol (BGP) was developed, ``built on experience gained with EGP''~\cite[p. 1]{rfc1105}. In his 2018 book, David Clark describes the shift to BGP as necessary to support a commercial network that would allow  %for 
different Internet Service Providers (ISPs) to have non-hierarchical relationships with one another (as opposed to EGP's more hierarchical structure, %that put 
with the ARPANET at the core). When discussing BGP's limitations, however, Clark raises the issue of its ``limited expressive power,'' providing different ADs (in this case ISPs) with minimal reachability information that could inform decisions about routing traffic through other ADs~\cite[pp. 244-245]{clark2018internet}.}

The network still maintained its original goal: being able to connect a large number of systems of different types of networks---to provide high-performance communication and interoperability among diverse hardware. But it now also had the %added 
goal of supporting ``Multiple organizations with mutual distrust and policy/legal restrictions''~\cite[pp. 4-5]{rfc1136}. The change in the network environment to one of mistrust unearthed critical architectural problems, as mistrust was fundamentally at odds with central assumptions in the Internet Protocol (IP), the network's core routing protocol. Routing ``determines the series of networks and gateways a packet will traverse in passing from the source to the destination''~\cite[p. 1]{rfc1102}. Embedded in the process is a notion of \emph{how} to determine the series  over which traffic should travel, which IP implemented by ``minimizing some measure of the route, such as delay''~\cite[p. 1]{rfc1102} and only promising to provide ``point-to-point best-effort data delivery''~\cite[p. 2]{rfc1633}. In only needing to satisfy the constraint of minimizing some cost, the selected route could hypothetically use any path to transport a packet from source to destination. This strategy would not be sufficient in an environment of mistrust: A packet traveling between two regions that trusted each other could potentially travel through an untrusted region while in transit. Point-to-point, best-effort service gave ADs no explicit control over how their packets were routed, instead locating best-effort routing decisions in the internals of the network~\cite[p. 1]{rfc1136}~\cite[p. 5]{rfc1125}. This left packet data---and any downstream effects that data might have on the destination's resources---at risk of interception or tampering.

As the only mechanism for transporting packets between ADs, best-effort routing was therefore in direct tension with the intra-AD management concerns %we have 
discussed in this section. For an AD's management goals to be met in the context of a distributed network, it would be important to add additional control mechanisms %of control 
``to select routes in order to restrict the use of network resources to certain classes of customers''~\cite[p. 1]{rfc1102}. It became clear that administrators needed to be able to set specific \emph{policies} for how to select routes---constraints for how data should traverse the network in order to conform with local resource usage goals. As we discuss in the next section, being able to support policies through network architecture would require fundamental changes. That is, changes for ``controlled network resource sharing and transit [would] require that policy enforcement be integrated into the routing protocols themselves and [could] \emph{not be left to network control mechanisms at the end points}''~\cite[p. 5, emphasis added]{rfc1125}. Policies would test the supposed fundamental commitment to End-to-End that had governed the network since its inception~\cite{saltzer1984endtoend}. Comprehensive accounting for resource usage would demand placing additional functionality in the network. Being able to connect was no longer enough; ADs needed policies to control how and when that connection occurred.

\section{Accounting as policy: From accounting to accountability} \label{sec:policy}

As we have shown in Section~\ref{sec:measurement}, in the early days of the ARPANET accounting was considered a ``supporting role compared to the core objective of forwarding data''~\cite[pp. 46-47]{clark2018internet}. But, as we discussed in Section~\ref{sec:management}, over the course of the 1980s and growth of the network, accounting, and its role in management, was directly implicated in the network's core objective of routing. Questions of routing started getting framed as questions of policy, and ARPANET engineers thus began to reckon with the notion that matters of technical design could not be separated from issues like who got to use the network and for what purpose. As David Clark noted back in 1989, ``Policy matters are driven by human concerns, and these have not turned out to be amenable to topological constraints, or indeed to constraints of almost any sort''~\cite[p. 2]{rfc1102}. Routing decisions could no longer just be about technical constraints, such as minimizing a cost metric like overall distance; they would also need to incorporate legal and political constraints so that routing could produce ``predictable, stable result[s] based on the desires of the administrator''~\cite[p. 2]{rfc1104}.

STS scholars have shown that technical and scientific work like this often involves such delineation of social and political aspects as existing beyond the scope of the engineers' work. Much of the literature produced by these scholars has focused on exploring the inherent embeddedness of technical decisions within social and political structures, clarifying that there is no real way to carve out the \emph{socio} from the \emph{technical}~\cite{bijker1987sts,latour1993modern,jasanoff2004states}. We can see this insight from STS play out in the network engineers' attempts to weed policy out of the network's design---to no avail. Even if policy was a ``human concern,'' the network still needed a technical architecture in order to implement it.

\subsection{Policy routing: An implementation for in-network control}\label{sec:policyrouting}

By the end of the 1980s, some engineers were arguing that the network needed a ``new generation of routing protocols'' that would allow each AD ``to independently express and enforce policies regarding the flow of packets to, from, and through its resources''~\cite[p. 1]{rfc1125}~\cite{estrin1989securityissues}.\footnote{And ultimately, even though accounting had become more expansive than the act of billing, policy remained inextricably tied to billing: ``The discussion of lost packets makes clear an important relationship between billing and policy. If a Policy Route takes packets through a region of known unreliability, the regions preceding it on the path may be quite unwilling to forgive the charges for packets which have successfully crossed their region, only to be lost further down the route. A billing policy is a way of asserting that one region wishes to divorce itself from the reliability behavior of another region. ... The use of a specific policy condition can make clear to the end user which [ADs] do not view themselves as interworking harmoniously''~\cite[p. 12]{rfc1102}.} These new architectural demands to support policies for routing control could not be separated from accounting. Accounting for resources was essential for controlling resources, since it was directly implicated in preventing, tracking, and correcting for unintended use~\cite[p. 9]{rfc1104}. The resulting proposed solution, ``Policy Routing'' (PR), was the architecture %al approach 
put forth to enable different communication policies between ADs in the network. 

A PR consisted of a sequence of gateways from source AD to destination AD; %(see Section~\ref{sec:management}); 
if such a route existed, then the policy associated with that PR was satisfied~\cite{breslau1990ad, rfc1102}.\footnote{There were various competing proposals on how exactly to implement policy routing. David Clark's RFC 1102 was among the most-cited, ultimately expanded upon by systems researcher Deborah Estrin~\cite{estrin1991policy}. Estrin collaborated on several different policy routing implementation schemes. All proposals, regardless of specifics, had to handle ``three design parameters: location of routing decision (i.e., predetermined at the source or hop-by-hop), algorithm used (i.e., link state or distance vector), and expression of policy in topology or in link status''~\cite[p. 231]{breslau1990ad}.} Such policies came in two flavors: access control and charging, %policies, 
frequently tied to quality of service (QoS)~\cite[p. 12]{rfc926}~\cite[p. 35]{rfc1009}.
Access control policies determined ``who [could] use resources and under what conditions''~\cite[p. 10]{rfc1125}. %These policies 
They enabled filtering out traffic considered ``administratively inappropriate''~\cite[p. 6]{rfc1126}, including blanket policies that defined ``users, applications, or hosts'' that could never ``be permitted to traverse certain segments of the network''~\cite[p. 9]{rfc1077},\footnote{While policy routing is not an architectural requirement within today's Internet, it remains an often-used enterprise solution that can be implemented in routers used by organizations and ISPs. Geoblocking, while similar in purpose to some access-based policies, is a different function for restricting traffic based on a user's geographical location. Rather than examining the internals of packet flows, traffic is blocked at the source using an IP address (a heuristic for geographic location, since IP addresses are generally assigned by country). Thus, as Clark notes, geoblocking is ``supported approximately'' on today's Internet; it is fairly easy to circumvent via using a VPN~\cite[p. 298]{clark2018internet}. For a detailed treatment of geoblocking, see~\citet{goldsmithwu2006geoblocking}. %In computer networking, 
The recent advent in software-defined networking (SDN) technology indicates that it is possible policy routing will experience renewed interest; SDN could allow for the implementation and enforcement of application-specific policies~\cite[p.  210]{clark2018internet}~\cite{greenberg2019cleanslatesdn}.} and finer-grained policies, which could allow a network manager to enforce traffic restrictions on a ``particular misbehaving host''~\cite[p. 34]{rfc1009} for a specific period of time. In contrast, charging policies could ``be based upon equity (`fairness') or upon inequity (`priority')''~\cite[p. 49]{rfc1009} %concerning 
of network traffic; they controlled the level of service guarantees for packet-forwarding, ranging from best-effort (with no service guarantees) to prioritized service, in which, for a premium, packets could jump to the head of the forwarding queue.\footnote{Even if the division of policy types %types of policies 
was clear at a high level, it %still 
remained unclear what a reasonable charging policy might be: e.g.,  %That is, 
even if %now 
it was clear \emph{who} was paying for a particular %quality of service
QoS, it was not clear ``that the services provided at the network layer [would] map well to the sorts of services that network consumers [were] willing to pay for. ... %''~\cite[p. 9]{rfc1104}. As noted in RFC 1104, ``
In the telephone network (as well as public data networks), users pay for End-to-End service and expect good quality service in terms of error rate and delay (and may be unwilling to pay for service that is viewed as unacceptable). In an internetworking environment, the heterogeneous administrative environment combined with the lack of End-to-End control may make this approach infeasible''~\cite[p. 9]{rfc1104}.} Policies enabling higher-quality service options would be useful for new types of %networked 
applications, like videoconferencing;\footnote{In the mid 1980s, new technical %software and hardware 
capabilities for end-node workstations presented the opportunity to develop entirely new classes of applications~\cite[p. 2]{rfc1633}: %Applications like 
Videoconferencing could be widely available, not just fodder for flashy %occasional 
demos, as in the late 1960s~\cite[pp. 138-142, Mother of All Demos]{bardini2001engelbart}. However, new workstation technology would not alone be sufficient to support such demanding applications; the network would also need to update its protocols. %in order to better support them. 
As mentioned above, %in Section~\ref{sec:policy}, 
best-effort routing is not well-suited to real-time applications like videoconferencing: packets get lost or dropped due to congestion; there are unpredictable periods of slow service due to packet queuing delays. Supporting additional qualities of service in network protocols would better accommodate these new hardware- and software-enabled applications.} such real-time applications are more sensitive to the disruptions common to best-effort service, including unpredictable network latency and delays. Additionally, on a saturated network, the ability to set policies to prioritize service could be useful to ADs that wanted to treat some types of traffic as more important than others. Network engineers were less concerned with discussing these specific policy choices, but rather wanted to ensure that the policy routing architecture was sufficiently flexible to implement a wide range of AD-specific policy requirements~\cite[p. 5]{rfc1102}~\cite[p. 4]{rfc1125}.

All of these policies, regardless of access- and QoS-related particulars, fundamentally concerned resource usage, and therefore ultimately implicated accounting. That is, while ``Network accounting [was] generally considered to be simply a step that leads to billing,'' by the late 1980s it was clear that accounting had much broader utility~\cite[p. 34]{rfc1077}, as these same ``records [could] also be used to determine usage patterns for the system''~\cite[p. 28]{rfc1244}. Irrespective of whether accounting records were used for billing, they contained useful information concerning resource usage, and therefore were essential for informing policy to control usage~\cite[p. 35]{rfc1077}~\cite[p. 11]{rfc1102}~\cite[p. 6]{rfc1125}.\footnote{Cost recovery and billing remained important considerations of policy: ``Almost all of the existing Internet has been paid for as capital purchase and provided to the users as a free good. There are limited examples of cost recovery, but these are based on an annual subscription fee rather than a charge related to the utilization. \textit{There is a growing body of opinion which says that accounting for usage, if not billing for it, is an important component of resource management. For this reason, tools for accounting and billing must be a central part of any policy mechanism}''~\cite[p. 11, emphasis added]{rfc1102}. And yet, just as the early conversations over billing in the 1970s indicated (Section~\ref{sec:billing}), setting an overarching billing policy is especially challenging in a distributed network, which was made even more difficult when different regions of the network clarified their individual administrative needs (Section~\ref{sec:management}): ``However, precisely because the administrative regions are autonomous, we cannot impose a uniform form of billing policy on all of the regions...The billing problem is thus a very complicated one, for the user would presumably desire to minimize the cost, in the context of the various outstanding conditions''~\cite[p. 11]{rfc1102}.} For example, records could be used in \emph{post hoc} audits to confirm that actual resource usage aligned with set access and QoS policies~\cite[pp. 49,66,77]{rfc1244}~\cite[p.11]{rfc1017}.\footnote{This same data, aside from setting and enforcing policy, also retained its overarching role of allowing ``network management personnel to determine the `flows' of data on the network, and the identification of bottlenecks in network resources,''~\cite[p. 11]{rfc1017} meaning it continued to support monitoring and diagnostics for operational reasons.} In particular, ``unusual accounting records [could] indicate unauthorized use of the system''~\cite[p. 28]{rfc1244}. If a pattern of ``malicious''~\cite[p. 11]{rfc1017} use or other of ``abuse (e.g., unauthorized use) develop[ed], an accounting system could track this and allow corrective action to be taken, by changing routing policy or imposing access control (blocking hosts or nets)''~\cite[p. 10]{rfc1104}. Routing policy tied together accounting information with the ``human concerns'' of how data moved around the network and introduced, into the architecture, considerations that conflicted with strict notions of efficiency.

\subsection{An unaccountable network: Enabling accountability required designing for accounting} \label{sec:unnaccountable}

This connection between accounting and %the 
enforcement of resource control policies brought about the necessary conditions to produce the first working definition of accountability in relation to the network. In RFC 1125, networking researcher Deborah Estrin made the interdependence between accounting, policy, and accountability unimpeachably clear:

%\begingroup
\addtolength\leftmargini{-.1in}
\begin{quote}
    \small{One way of reducing the compromise of autonomy associated with interconnection is to implement mechanisms that assure accountability for resources used. Accountability may be enforced a priori, e.g., access control mechanisms applied before resource usage is permitted. Alternatively, accountability may be enforced after the fact, e.g., record keeping or metering that supports detection and provides evidence to third parties (i.e., non-repudiation). Accountability mechanisms can also be used to provide feedback to users as to consumption of resources. ... [I]t becomes more appropriate to have resource usage visible to users, whether or not actual charging for usage takes place~\cite[p. 6]{rfc1125}.}
\end{quote}
%\endgroup

\noindent In short, achieving accountability in the network meant being able to implement policies for dealing with resource misuse. While some policies emphasized prevention and others concerned identifying, isolating, and mitigating misuse after-the-fact, %it had already occurred, 
all policies were ultimately dependent on accounting records~\cite[p. 9]{rfc1104}.\footnote{RFC 1104 tries to distinguish the function of accounting from the function of policy-based routing. In the process, it becomes clear that, while accounting's role is more expansive than its implications in policy-based routing, policy-based routing entirely depends on accounting: ``Accounting vs. Policy Based Routing: Quite often Accounting and Policy Based Routing are discussed together. While the application of both Accounting and Policy Based Routing is to control access to scarce network resources, these are separate (but related) issues. The chief difference between Accounting and Policy Based Routing is that Accounting combines history information with policy information to track network usage for various purposes. Accounting information may in turn drive policy mechanisms (for instance, one could imagine a policy limiting a certain organization to a fixed aggregate percentage of dynamically shared bandwidth). Conversely, policy information may affect accounting issues''~\cite[p. 9]{rfc1104}.}

In showing the importance of accounting for enabling accountability, Estrin's RFC 1125 captures the fundamental argument that we have taken up in this paper: The tensions that come up in a resource-sharing, networked computing environment (Sections~\ref{sec:billing} \&~\ref{sec:measurement}), in which there is no global authority (Section~\ref{sec:management}), ultimately reflect tensions concerning accountability and autonomy. While we have shown throughout this paper that issues of resource management were never not contentious, it was at this point in time that the network architects and engineers found accounting-related issues sufficiently important to be considered at the level of architecture, instead of an annoyance. In Estrin's words, ``the lack of global authority, the need to support network resource sharing as well as network interconnection, the complex and dynamic mapping of users to ADs and rights, and the need for accountability across ADs, are characteristics of inter-AD communications which must be taken into account in the design of both policies and supporting technical mechanisms''; ``it would be inexcusable to ignore resource control requirements and not to pay careful attention to their specification''~\cite[pp. 6,7]{rfc1125}.

Rather than accounting not meeting the parsimony requirements of End-to-End, accounting's necessary role in accountability meant that the engineers promoting a policy routing architecture---including, notably, End-to-End~\cite{saltzer1984endtoend} author and network architect David Clark~\cite{rfc1102}---were willing to consider accounting features as sufficiently
fundamental to incorporate within the network's foundational routing protocol. This would require placing mechanisms for accounting in the network, instead of just at the end nodes. This perspective marked a significant shift from the early days of the ARPANET, well-characterized in Bob Kahn's RFC 136 (Section~\ref{sec:measurement}). Before, accounting was discounted as not being a part of the network's core architecture, with respect to the End-to-End principle: It was separated from the function of routing and was not a ``performance enhancement''~\cite{saltzer1984endtoend, rfc136}, and therefore was not considered a candidate feature of the network’s essential architecture. By 1988, recognizing accounting's necessity for accountability placed accountability directly at odds with this earlier interpretation of End-to-End. To resolve this tension by choosing to acknowledge accountability as an essential feature, the network's engineers would be forced to re-imagine the meaning and primacy of End-to-End. For an accountable network, the next evolution of its architecture would need to be shaped by a new, competing ideology.

\section{Conclusion} \label{sec:conclusion}

In this paper, we have traced the changing meaning of the term ``accounting'' among the research community which developed the ARPANET and early Internet, from the late 1960s and through the end of the 1980s, with a particular focus on the RFC corpus from this time period. We have paid attention both to how accounting was and was not considered part of the set of research problems facing network engineers, and to how the meaning of ``accounting'' shifted in relation to the changing environment of the network's deployment and institutional context. We characterized four notions of accounting within ARPANET and early Internet RFCs---accounting as \emph{billing}, accounting as \emph{measurement}, accounting as \emph{management}, and accounting as \emph{policy}---and argued that the different meanings of accounting and the stakes of debates concerning each theme provide an emerging articulation of accountability. Our analysis places accounting and its administrative associations squarely within the domain of both deep technical questions about the network and the political expectations of accountable technical systems. 

This analysis resonates beyond the historiographical considerations of the Internet's development and the contemporary policy issues of its governance. It demonstrates that accountability is not a pre-given quality of a system that can be designed from the get-go, but rather is a set of emergent attributes that are negotiated throughout the design process---a process that itself is never fully settled in deployed systems like the Internet. From this starting point, we can see why technical systems that are now identified as unaccountable cannot be remedied simply by calls to implement greater accountability, which also run the risk of being mistakenly equated with a call to record everything as an audit best practice. Rather, taking stock of accountability as the result of complex negotiations, in the ways we have done above, means first and foremost recognizing whose interests are accounted for in a given accountability regime. 

We close by offering three insights from this discussion that have bearing on research concerning accountability and complex technical systems, which together serve as a cautionary tale for the design of contemporary computer systems. First, the core design question of the story above is about resource sharing and allocation, a motivating problem in many computing applications today, especially in machine learning. The non-trivial questions involved in designing mechanisms to account for resource use in a distributed system are constitutive of the possibility of creating accountability. It is thus worth interrogating how mechanisms for accounting may facilitate or obscure accountability. Second, we argued that developing accounting mechanisms was routinely deferred, not only because of the foundational tensions involved in developing accounting schemes, but because of a dynamic of discounting its significance. Accounting used the administrative language around issues of delineating different actors' autonomy and negotiating trust, which created a policy of prioritization, allowing such issues to be dismissed as ``operational,'' beyond the core set of research concerns, even though it was clear that distributed accounting for resource sharing was a fundamental challenge in the field of networking. Finally, by taking an institutional approach to the history of technical objects, we argued that the broader context in which technological systems develop shapes the ways accountability is defined and implemented (or ignored). Taking the administrative needs of this setting seriously, rather than casting them aside as overhead concerns, would help clarify the necessary infrastructure to support concrete notions of accountability in complex technical systems. Even within the same technical system---%for example, 
the Internet---we see how over time institutional changes led to shifts in understandings of accounting and the demands of accountability. Designing a definitive ``one-size-fits-all'' technical spec for accountability, thus, would offer an abstract solution to a set of complex questions that ought to be addressed within specific institutional contexts.

\begin{acks}
A. Feder Cooper is supported by the Artificial Intelligence Policy and Practice initiative at Cornell University, the Digital Life Initiative at Cornell Tech, and the John D. and Catherine T. MacArthur Foundation. The authors would like to thank the Artificial Intelligence, Policy, and Practice initiative, the Fidelity, Authenticity, and Knowability in Electronic media Lab, and the Qualitative and Interpretive Research Institute at Cornell, FAccT’s anonymous reviewers, and the following individuals for feedback on earlier versions of this work: Harry Auster, Salonee Bhaman, Solon Barocas, Madiha Zahrah Choksi, Fernando Delgado, Abigail Z. Jacobs, Devin Kennedy, Jon Kleinberg, Karen Levy, Pegah Moradi, and Malte Ziewitz.
\end{acks}

%\balance
\bibliographystyle{ACM-Reference-Format}
\bibliography{references}

%\newpage
%\input{sections/appendix/appendix}

\appendix

\section{The ``Research'' vs. ``Service'' Distinction} \label{app:sec:researchservice}

It is important to note that Bressler~\cite{rfc487} and Pogran~\cite{rfc501} were not alone in conceiving of rationales for circumventing accounting; punting on both implementing and performing accounting for certain aspects of network use was a common theme in the early ARPANET. Most notably, there was an overarching attempt to distinguish categories of work on the network, such that some preferred types of work could be ``free''---that is, have the bill covered by ARPA. In response to this desire, a binary distinction emerged at the site level---``Research Centers vs. Service Centers''~\cite[p.1]{rfc77}; or, ``free but limited access research sites'' used strictly for experimental purposes and ``billing sites''~\cite[p. 3, p. 18]{rfc82}. Early Internet pioneer Jon Postel further refined this classification in relation to hardware and access patterns: ``The Service Centers tend to have big machines, lots of users, and accounting problems; while the Research Centers tend to have specialized hardware, a small number of users, and no accounting at all''~\cite[p.1]{rfc77}. Even with these definitions, it is not entirely clear what ``service'' meant in terms of function, aside from the common need for accounting in order to support site use. The word ``service'' remains simultaneously vague and overloaded, both during the 1970s when ARPANET engineers were teasing out research/service distinction and with respect to contemporary use. For example, in his 2018 book, David Clark notes that ``service'' is a term that has repeatedly confused him, and that he makes sense of it by reducing it to the following: ``A service is something that you sell; it is how you make money''~\cite[p. 311]{clark2018internet}.

One can attempt to elicit a far more precise understanding of service from Service Center examples, which included the Network Information Center (NIC) at SRI, Multics at MIT~\cite[p. 13]{rfc101}, UCSB's Simple-Minded File System (SMFS, which provided a secondary storage node that anyone on the ARPANET could pay to use; it had limited availability and was not intended to become the storage node for the whole network)~\cite{rfc122}, and UCSD's FTP service site, which had to bill for usage in order to support itself (each FTP file transfer was billed separately in the accounting system, based on lower-level accounting for processor, I/O,  and core usage, and, if used, external storage tapes)~\cite{rfc532}.  

However, these varied examples of site functions, particularly the inclusion of FTP, the costs of which were debated in unique detail (Section~\ref{sec:freeftp}), indicate that what constituted the distinction between research and service was contentious and unclear.\footnote{This contention is explicitly addressed as a difference in ``orientation'' in RFC 231: ``In the network at large, with our research orientation, personnel tend to have a different approach to computing than that required by a service bureau.'' Service Centers were believed to be subject to ``market-oriented requirements'' to rate-limit use, while Research Centers were free from such forces~\cite[p. 4]{rfc231}.} Moreover, it was additionally not trivial to consider how billing should be handled when research and service sites resource share. Notably, Larry Roberts acknowledged this, but did not provide a clear idea on how to resolve it: ``What happens when a research site talks to a billing site? I think it is do-able''~\cite[p. 3]{rfc82}. The UCSD FTP site, at least by 1973, had to account for usage and issue bills in order to support itself (it is unclear if this site had an option to use a FREE account for research purposes). However, as the RFCs debating ``free'' file transfer show, this need for accounting did not extend to all FTP usage, whether it was hosted at a service site or not---at least not immediately~\cite{rfc487, rfc501}. 

That is, even if accounting was not necessary at research sites to start, it was acknowledged that, as more users joined the ARPANET and wanted access to limited resources, it would eventually be necessary to account and bill for research usage, as well. For example, as early as RFC 82, Douglas Engelbart was recorded as saying that SRI will eventually have to bill as more users come online: ``A system will exist in Spring 1971, to allow an agent to insert into a catalog. The dialogue that goes on will determine which way the data base grows. We are pretty sure that eventually SRI will have to charge because of many potential users not at primary sites seeking limit[sic] resources. ... Each site is registered. Any person who gets in on a site's account has its access. We won't worry about accounting until saturation occurs. We would like to encourage use of the agent system to create and use a survey of resources at each site,''~\cite[p. 7]{rfc82} which included SRI's research theorem prover tools~\cite[p. 548]{robertswessler1970resourcesharing}. In fact, by 1987, the interagency research Internet proposal made it clear that accounting would absolutely be necessary for research~\cite{rfc1015, rfc1017}.

\section{The Decline of the Ideal of Resource Sharing} \label{app:sec:abbateusability}

Janet Abbate argues that, as the network began to spread to more ARPA-contractor sites, the ``demand for remote resources fell,'' and that ``many sites rich in computing resources seemed to be looking for users''~\cite[p. 104]{abbate1999inventing}. Abbate credits this ``decline of the ideal of resource sharing'' (where this ideal was the vision of the ARPANET articulated in ~\citet{roberts1967arpanetproposal}) to severe usability issues in the ARPANET. Throughout Chapter 3, she discusses the practical difficulties of using the network---even with an action as simple as finding a particular resource, as the network lacked appropriate search tools~\cite[p. 86]{abbate1999inventing}.\footnote{See also RFC 531, concerning ARPANET usability problems, the creation of a resource notebook to improve these problems, and ultimately the additional gaps the resource notebook highlighted in relation to documentation reliability issues~\cite{rfc531}.} 

Even if users got past the initial hurdle of finding a resource, there remained additional steps in order to access it. Abbate acknowledges that this in part had to do with accounting, but discusses the issue as one of usability: The user would have to contact an administrator to set up an account on the remote host in order to access it, which required finding and contacting the appropriate person at the remote site; if the remote site wanted to charge for usage, the user then also usually had to initiate a purchase order at their local institution. Only then could they access the resource, at which point it often remained a challenge to figure out how the resource was supposed to be used~\cite[pp. 87-88]{abbate1999inventing}. This lack of usability, which reflected both technical and administrative issues, ideally should not have been a relevant concern for individual users. However, it was a particularly challenging obstacle for novice users, and thus became a bottleneck in the network's ability to grow and reach saturation. Abbate therefore reasons that the ARPANET became a technology seeking an appropriate application and an interested user base, as it fell short of its goal to facilitate resource sharing. The ARPANET needed to experience a fundamental shift in ``identity and purpose'' if it was to be a useful technology~\cite[p. 109]{abbate1999inventing}.\footnote{For more on evaluating the ``friendliness'' of the network, see RFC 369 ~\cite{rfc369}; see RFC 451~\cite{rfc451} and RFC 666~\cite{rfc666} for defining the unified user level protocol (UULP)---a proposed solution to ARPANET usability issues, which suggested a command language for ``user convenience,'' ``'resource sharing','' ``economy of mechanism,'' ``front-ending...onto existing commands,'' ``accounting and authorization,'' and ``process-process functions''~\cite[pp. 1-2]{rfc666}. In other words, UULP was an attempt to come up with a single protocol that would help with network usability functions.}

Abbate writes that this shift ultimately occurred when the ARPANET found such a ``smash hit'' application in email~\cite[p. 106]{abbate1999inventing}. In contrast to the technical and administrative usability issues that plagued resource sharing, email was very simple to use; it connected users at remote sites, but was an application that users could access locally. Similarly, local area networks (LANs) also became a popular, unexpected use of the ARPANET at this time, as LANs did not have the same usability issues as more long-range, distributed resource sharing. By 1975, with sites like USC and SRI using the ARPANET as for LANs, 30\% of traffic on the network was intra-node, as opposed to inter-node~\cite[p. 94]{abbate1999inventing}. In short, Abbate argues that, while resource sharing struggled to find users, these two ARPANET uses were successes that validated the utility of the network.

Unpacking Abbate's focus on usability, we argue that it is possible to understand the ``decline of the ideal of resource sharing'' and its attendant challenges to the would-be user in terms of the lack of appropriate, fleshed-out accounting mechanisms. While usability was certainly a relevant factor concerning the (lack of) ease of adoption of the ARPANET for resource sharing, it should not be conflated with a lack of desire to resource share altogether. In fact, the debate over ``free'' file transfer (Section~\ref{sec:freeftp}) and the attempt to classify different sites as research or service centers is evidence that ARPANET users wanted to resource share (with services like UCSB's SMFS and UCSD's FTP node being sufficiently utilized to require accounting to recoup costs, see Appendix \ref{app:sec:researchservice}). Instead, as we discussed in Sections~\ref{sec:billing} and ~\ref{sec:measurement}, accounting was (and remains) an extremely challenging technical and administrative problem, which the early ARPANET architects did not have an appetite to address. As a result, accounting did not become a first-order feature of the network, and was instead a messy patchwork of non-interoperable, ill-defined systems that made recovering the costs of resource sharing intractable.

From this perspective, it is possible to recast Abbate's examples of unexpected successes of usability---the ``smash hit'' of email and the proliferation of LANs---as successes owing to their independence from distributing accounting. Email leveraged the distributed network, but it did not have the same distributed accounting problems as resource sharing. Aside from using gateway nodes for routing email to its final destination, email (at least its 1970s iteration) was a local application, using local CPU and local storage. If accounting needed to be done, it could be handled locally at the site level, using local administrative procedures (if there were any), such as those used for time-sharing~\cite[p. 4]{rfc33}~\cite[concerning UCLA's internal billing for email]{rfc555}. In other words, it is possible to view email as a success not just because of ease of use, but also because it was possible to treat email as ``free'' in a way similar to the initial assumptions about the negligible costs of ``free'' file transfer, in which local billing could be used for local usage or billing could be punted back to ARPA. Of course, just like ``free'' file transfer, email was not literally free. Email eventually became quite costly, particularly when sending spam became a growing practice.\footnote{For a detailed treatment on the history of Internet spam, see \citet{brunton2013spam}.} Spam put a strain on receivers' local resource usage, which meant that it, too, became an accounting problem~\cite{rfc2635}.\footnote{RFC 2635, ``Don't Spew: A Set of Guidelines for Mass Unsolicited Mailings and Postings (spam*),'' discusses the costs of spam via a comparison with physical mail. It notes that it is easier to send email, so the scale of junk email is much greater. The costs are also quite different: It costs the sender very little to send spam; ``the recipient bears the majority of the cost''~\cite[pp. 3-4]{rfc2635}. The RFC thus calls spam ``unethical behavior''~\cite[p. 3]{rfc2635}, and goes so far as to compare it to a seizure of private property, since it eats up local resources---a ``theft of service''~\cite[p. 4]{rfc2635}} Similarly, LANs were intra-node, and also represented a problem of local accounting, as opposed to distributed accounting.

\section{Computational Methods} \label{app:sec:methods}

In this appendix, we document our procedure for identifying RFCs related to our project. The accompanying code can be found at \href{https://github.com/pasta41/rfc-scraper}{\texttt{https://github.com/pasta41/rfc-scraper}}.

\vspace{.1cm}\noindent\textbf{Our \href{https://github.com/pasta41/rfc-scraper}{\texttt{rfc-scraper}} tool.} We developed a tool to scrape the \href{https://www.rfc-editor.org/rfc/}{\texttt{RFC Editor}}, which pulls down the \texttt{.txt} version of each RFC and associated \href{https://www.rfc-editor.org/rfc-index.html}{metadata}. We wrote a separate script to filter and map the above data to identify RFCs. The filtering capability is simple: The script takes as an argument a search term to \texttt{grep} for (e.g., ``account'', treated as a prefix and case insensitive); if there is a match in an RFC, the RFC ID is mapped to its associated metadata (author, title, working groups, etc.) and the results are saved to a \texttt{.csv} file. Separately, each matched instance in each RFC is printed to a \texttt{.html} file (with 5 leading and trailing lines of surrounding context, and each matched word highlighted in color and emphasized).

For more detailed documentation on this code, please refer to \href{https://github.com/pasta41/rfc-scraper}{\texttt{https://github.com/pasta41/rfc-scraper}}. The \href{https://github.com/pasta41/rfc-scraper/blob/main/README.md}{\texttt{README}} in the repository has the most up-to-date information related to the required python dependencies (including bash scripting utilities called within python for extra efficiency), and the ANSI HTML adapter package used to develop the color-coded \texttt{.html} search results files. We intend to update this tool with more sophisticated search functionality so that others can use it in the future for additional RFC-related research (e.g., filtering by time range, working group, author;  other data manipulation functions).

\vspace{.1cm}\noindent\textbf{Search terms and verification process.} We ran the \href{https://github.com/pasta41/rfc-scraper}{\texttt{rfc-scraper}} tool on several search terms related to our project purpose:

\begin{itemize}
    \item ``account'', which matches \texttt{account*} and served as our superset search for all accounting terms
    \item ``accounting'', which matches a subset \texttt{account*}
    \item ``accountable'', which matches a subset of \texttt{account*}
    \item ``accountability'', which matches a subset of \texttt{account*}
    \item ``time-shar'', which matches \texttt{time-shar*} and therefore includes \texttt{time-share}, \texttt{time-sharing}, and similar terms
    \item ``survivabl'', which matches \texttt{survivable} and \texttt{survivability}
\end{itemize}

We then manually read the \texttt{.html} files, which include leading and trailing text surrounding matches, to determine which RFCs we should read in more detail for our project. Of the 9085 RFCs published at the time of running the tool, this process enabled us to identify a subset of 136 RFCs to read in full. 19 of the 136 were false positives (i.e., they were not ultimately relevant). During detailed reading of the 136 RFCs, we further identified 12 RFCs that our search terms missed, which were relevant to our project. We found these RFCs through references in both the RFCs our tool selected and in the secondary literature we cite. We include our RFC tracker code book in the \href{https://github.com/pasta41/rfc-scraper}{\texttt{rfc-scraper}} repository. 
As an aside, through this process we learned that our search tool bears a coincidental resemblance to (though is much simpler than) the search-and-filter tools built to identify time datatype issues related to the Y2K in Internet protocols~\cite{rfc2626}.

\end{document}